\documentclass[a4paper,12pt]{article}
\usepackage[pctex32]{graphics}
\usepackage{amssymb,amsmath}

\begin{document}

\topmargin 0pt \oddsidemargin 0mm
\newcommand{\be}{\begin{equation}}
\newcommand{\ee}{\end{equation}}
\newcommand{\p}{\partial}

\begin{titlepage}

\vspace{5mm}
\begin{center}
{\Large \bf Entropy of an extremal regular black hole }

\vspace{12mm}

{\large   Yun Soo Myung$^{\rm a,}$\footnote{e-mail
 address: ysmyung@inje.ac.kr},
 Yong-Wan Kim $^{\rm a,}$\footnote{e-mail
 address: ywkim65@naver.com},
and Young-Jai Park$^{\rm c,}$\footnote{e-mail
 address: yjpark@sogang.ac.kr}}
 \\

\vspace{10mm}

{\em $^{\rm a}$Institute of Mathematical Science and School of
Computer Aided Science, \\Inje University, Gimhae 621-749, Korea \\}
{\em $^{\rm b}$Department of Physics  and Center for Quantum
Spacetime,\\ Sogang University, Seoul 121-742, Korea}
\end{center}

\vspace{5mm}

\centerline{{\bf{Abstract}}}

\vspace{5mm}

We introduce a  magnetically charged extremal regular black hole in
the coupled system of Einstein gravity and nonlinear
electrodynamics. Its near horizon geometry is given by $AdS_2\times
S^2$. It turns out that the entropy function approach does not
automatically lead to a correct entropy of the Bekenstein-Hawking
entropy. This contrasts to the case of the extremal
Reissner-Norstr\"om black hole in the Einstein-Maxwell theory. We
conclude that the entropy function approach does not work for a
magnetically charged extremal regular black hole without
singularity, because of the nonlinearity of the entropy function.

\vspace{3mm}

\noindent PACS numbers: 04.70.Dy, 04.70.Bw, 04.20.Dw, 04.20.Jb  \\
\noindent Keywords: Regular black hole; Nonlinear electrodynamics;
          Entropy
\end{titlepage}


\section{Introduction}

Growing interest in the extremal  black holes is motivated by their
unusual and not fully understood nature. The problems of entropy,
semiclassical configurations, interactions with matter, or
information paradox have not been resolved yet. Moreover, the
near-horizon region is also of interest apart from their whole
structure and behavior. Indeed, one can generate new exact
solutions~\cite{gp} applying appropriate limiting procedure in the
geometry of the extremal and near extremal black holes. In
particular, of the equal importance is the question of the nature of
singularities that reside in the centers of most black holes hidden
to an external observer. Regular black holes (RBHs) have been
considered, dating back to Bardeen~\cite{BAR1}, for avoiding the
curvature singularity beyond the event horizon~\cite{RBH1}. Their
causal structures are similar to the Reissner-Nordstr\"{o}m black
hole with the singularity replaced by de Sitter
space-time~\cite{Dymn1}. Hayward has discussed the formation and
evaporation process of a RBH~\cite{HAY,0702145}. A more rigorous
treatment of the evaporation process was carried out for the
renormalization group improved black hole~\cite{REU}. The
noncommutativity also provides another RBH: noncommutative black
hole~\cite{0606051}.

Among various RBHs known to date, especially intriguing  black holes
are from the known action of Einstein gravity and nonlinear
electrodynamics. The solutions to the coupled equations were found
by Ay\'on--Beato and Garc\'{\i}a~\cite{9911174} and by
Bronnikov~\cite{0006014}.  The latter describes a magnetically
charged black hole, and provides an interesting example of the
system that could be both regular and extremal. Also its simplicity
allows exact treatment such that the location of the horizons can be
expressed in terms of the Lambert functions~\cite{0010097}.
Moreover, Matyjasek investigated the magnetically charged extremal
RBH with  the near horizon geometry of $AdS_{2}\times S^{2}$ and its
relation with the exact solutions of the Einstein field
equations~\cite{0403109,0606185}. Only this type of  RBHs can be
employed to test whether the entropy function approach is or not
suitable for obtaining the entropy of the extremal RBHs.

On the other hand, string theory suggests that higher curvature
terms can be added to the Einstein gravity~\cite{Zwiebach:1985uq}.
Black holes in higher-curvature gravity \cite{Callan:1988hs} were
extensively studied during two past decades culminating in recent
spectacular progress in the microscopic string calculations of the
black hole entropy. For a review, see \cite{deWit:2005ya}. In
theories with higher curvature corrections, classical entropy
deviates from the Bekenstein-Hawking value and can be calculated
using Wald's formalism \cite{9307038}. Remarkably, it still exhibits
exact agreement with string theory quantum predictions at the
corresponding level, both in the BPS
\cite{Behrndt:1998eq,Dabholkar:2004dq} and non-BPS
\cite{Goldstein:2005hq} cases. In some supersymmetric models with
higher curvature terms, exact classical solutions for static black
holes were obtained \cite{Dabholkar:2004dq}. Recently, Sen has
proposed a so-called ``entropy function'' method for calculating the
entropy of $n$-dimensional extremal singular black holes, which is
effective even for the presence of higher curvature terms. Here the
extremal black holes are characterized by the near horizon geometry
$AdS_{2}\times S^{n-2}$ and corresponding isometry~\cite{0506177}.
It states that the entropy of such kind of extremal black holes can
be obtained by extremizing the ``entropy function'' with respect to
some moduli on the horizon. This method does not depend  on
supersymmetry and has been applied to many solutions in supergravity
theory. These are extremal black holes in higher dimensions,
rotating black holes and various non-supersymmetric black
holes~\cite{SSen1,0604028}.

In this paper we consider a magnetically charged  RBH with near
horizon geometry $AdS_2\times S^2$ in the coupled system of the
Einstein gravity and nonlinear electrodynamics
\cite{0403109,0606185}. The solution is parameterized by two
integration constants and a free parameter. Using the boundary
condition at infinity, the integration constants are related to
Arnowitt-Deser-Misner (ADM) mass $M$ and magnetic charge $Q$,
while the free parameter $a$ is adjusted to make the resultant
line element regular at the center. Here we put special emphasis
on its extremal configuration because it has the same near horizon
geometry $AdS_2\times S^2$ of the extremal Reissner-Nordstr\"om
black hole ($a=0$ limit), but it is regular inside the event
horizon. In this work, we investigate whether the entropy function
approach does work for deriving the entropy of a magnetically
charged extremal regular black hole without singularity.

As a result, we show that the entropy function approach proposed
by Sen does not lead to a correct form  of the Bekenstein-Hawking
entropy of an extremal RBH. However, using the generalized entropy
formula based on Wald's Noether charge formalism~\cite{CC}, we
find the correct entropy.

\section{Magnetically charged RBH}

We briefly recapitulate a magnetically charged extremal RBH with the
special emphasis put on the near horizon geometry ${\rm
AdS}_{2}\times {\rm S}^{2}$ and its relation with the exact
solutions of the Einstein  equations~\cite{0403109,0606185}. Let us
begin with the following action describing the Einstein
gravity-nonlinear electrodynamics
\begin{equation}
S\,=\int d^4x \sqrt{-g}{\cal L}=\frac{1}{16\pi}\int d^{4}x \sqrt{-g}
\left[R\,-\,{\cal L}(B) \right].
                                         \label{action}
\end{equation}
Here ${\cal L}(B)$ is a functional of $B= F_{\mu\nu}F^{\mu\nu} $
defined by
\begin{equation}  \label{lagr}
 {\cal L}(B)=B \cosh^{-2}\left[a \left(\frac{B}{2}\right)^{1/4} \right],
 \end{equation}
where the free parameter $a$ will be adjusted to guarantee
regularity at the center. In the limit of $a\to 0$, we recover the
Einstein-Maxwell theory in favor of the Reissner-Nordstr\"om black
hole.
 First, the tensor field $F_{\mu\nu}$ satisfies equations
\begin{equation}
\label{maxwell1} \nabla _{\mu}\left( \frac{d{\cal L}(B)}{dB}
F^{\mu\nu}\right) =0,
\end{equation}
\begin{equation}
\label{maxwell2} \nabla _{\mu}\,^{\ast }F^{\mu\nu}=0,
\end{equation}
where the asterisk denotes the Hodge duality. Then, differentiating
the action $S$ with respect to the metric tensor $g_{\mu\nu}$ leads
to
\begin{equation} \label{EEQ}
R_{\mu\nu}-\frac{1}{2} g_{\mu\nu}R = 8\pi T_{\mu\nu}
\end{equation}
with the stress-energy tensor
\begin{equation}
T_{\mu\nu}=\frac{1}{4\pi }\left( \frac{d {\cal L}\left( B\right)
}{dB}F_{\rho \mu}F^{\rho}_{\nu}-\frac{1}{4}g_{\mu\nu} {\cal L}\left(
B\right) \right).
\end{equation}
Considering a static and spherically symmetric configuration, the
metric can be described by the line element
\begin{equation} \label{lineel}
 ds^{2}\,=\,-G(r) dt^{2}+ \frac{1}{G(r)}dr^{2}\,+
 \,r^2 \left(d\theta^2+\sin^2\theta d\phi^2\right)
\end{equation}
with the metric function
\begin{equation}
G(r)\,=\,1\,-\,\frac{2 m(r)}{r}.
 \end{equation}
Here, $m(r)$ is the mass distribution function. Solving the full
Einstein equation (\ref{EEQ}) leads to the mass distribution
\begin{equation} \label{quadrature1}
m(r)\,=\,\frac{1}{4}\int^r  {\cal L}[B(r')]r'^{2} dr'\, + C,
\end{equation}
where $C$ is an integration constant. In order to determine $m(r)$,
we choose the purely magnetic configuration as follows
\begin{equation}
F_{\theta\phi} = Q \sin\theta \to B=\frac{2Q^{2}}{r^{4}}.
\end{equation}
Hereafter we assume that $Q>0$ for simplicity. Considering the
condition for the ADM mass at infinity ($m(\infty)\,=\,M=C$), the
mass function takes the form
\begin{equation}
m(r)\,=M-\frac{Q^{3/2}}{2a} \tanh\left(\frac{aQ^{1/2}}{r} \right).
\end{equation}
Finally, setting $a\,=\,Q^{3/2}/2M$ determines the metric function
completely as
\begin{equation}
G(r)\,=\,1\,-\,\frac{2 M}{r}\left(1\,-\,\tanh\frac{Q^{2}}{2Mr}
\right). \label{Gr}
 \end{equation}
At this stage we note that the form of metric function $G(r)$ is
obtained  when using the mass distribution (\ref{quadrature1}) and
boundary condition. However, we will show that considering the
attractor equations (\ref{e33}) and (\ref{e332}) which hold in the
near horizon region only, one could not determine $G(r)$.  Also,
it is important to know that $G(r)$ is regular as $r \to 0$, in
contrast to the Reissner-Nordstr\"om case ($a=0$ limit) where its
metric function of $1-2M/r+Q^2/r^2$ diverges as $r^{-2}$ in that
limit. In this sense, the regularity is understood here as the
regularity of line element rather than the regularity of
spacetime.

In order to find the location $r=r_\pm$ of event horizon from
$G(r)=0$, we use the Lambert functions $W_i (\xi)$ defined by the
general formula $e^{W(\xi)}W(\xi)=\xi$ \cite{0403109}. Here
$W_0(\xi)$ and $W_{-1}(\xi)$ have real branches, as is shown in
Fig. 1a. Their values at branch point $\xi=-1/e$ are the same as
$W_{0}(-1/e)=W_{-1}(-1/e)=-1$.
 Here we set
$W_{0}(1/e) \equiv w_0$ because the Lambert function at $\xi=1/e$
plays an important role in finding the location $r=r_{ext}$ of
degenerate horizon for an extremal RBH. For simplicity, let us
introduce a reduced radial coordinate $x=r/M$ and a charge-to-mass
ratio $q=Q/M$ to find the outer $x_+$ and inner $x_-$ horizons  as
\begin{equation}
x_+=-\frac{q^2}{W_0(-\frac{q^2e^{q^2/4}}{4})-q^2/4},
~~x_-=-\frac{q^2}{W_{-1}(-\frac{q^2e^{q^2/4}}{4})-q^2/4}.
\end{equation}
Especially for $q=q_{ext}=2\sqrt{w_0}$ when
$(q_{ext}^2/4)e^{q_{ext}^2/4}=1/e=w_0e^{w_0}$, the two horizons
$r_+$ and $r_-$ merge into a degenerate event horizon
\footnote{For the Reissner-Norstr\"om black hole ($a=0$ limit), we
have the outer $r_+$ and inner $r_-$ horizon at $r_{\pm}=M \pm
\sqrt{M^2-Q^2}$. Further, its degenerate event horizon appears at
$r_{ext}=M=Q$. In terms of $x_\pm=r_\pm/M$ and $q=Q/M$, we have
$x_{\pm}=1\pm \sqrt{1-q^2}$. In the case of extremal black hole
($q^2_{ext}=1$),
 one has $x_{\pm}=x_{ext}=1$. Its entropy is given by
 $S^{RN}_{BH}=\pi M^2 x^2_{ext}=\pi Q^2$.} at
\begin{equation} \label{locebh}
x_{ext}=\frac{4q^2_{ext}}{4+ q^2_{ext}}=\frac{4w_0}{1+w_0}.
\end{equation}
This is shown in Fig. 1b. Alternatively, in addition to $G(r)=0$,
requiring a further condition
\begin{equation} \label{cebhole} G'(r)=0, \end{equation} one arrives at the same
location of degenerate horizon as in Eq. (\ref{locebh}). Here $'$
denotes the derivative with respect to $r$.
 For $q>q_{ext}$, there is no horizon.
\begin{figure}[t!]
   \centering
   \includegraphics{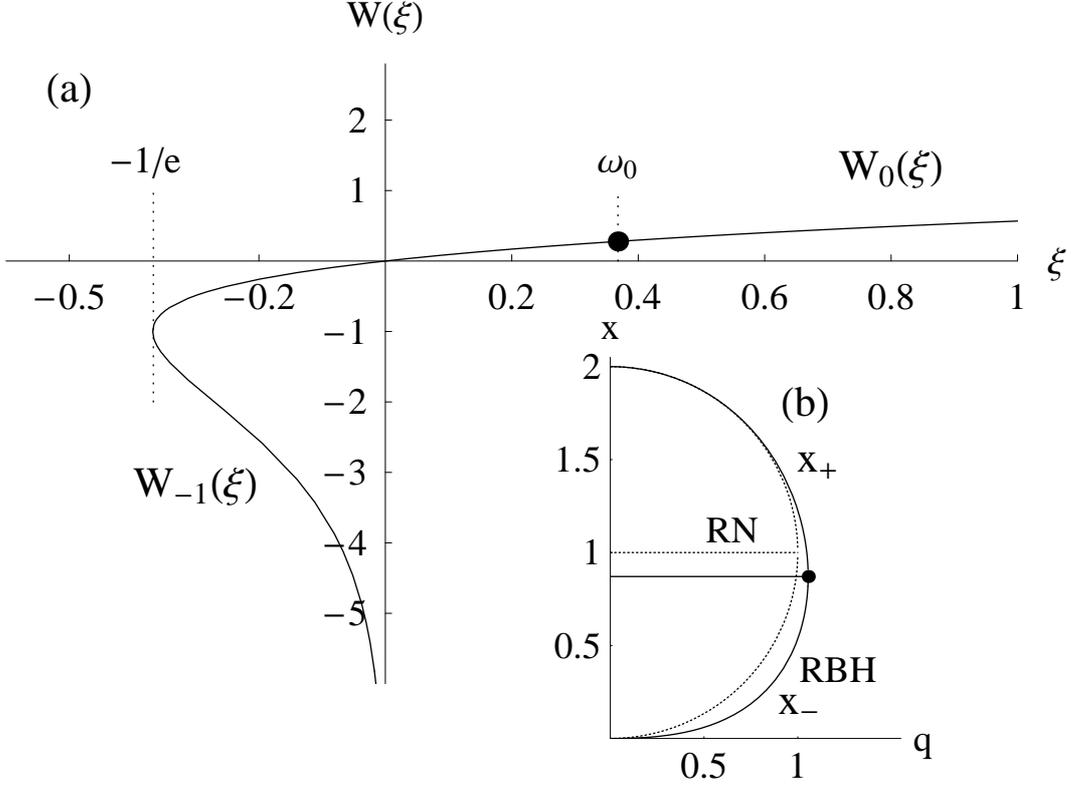}
\caption{(a) Two real branches of the Lambert function $W_0(\xi)$
(upper curve) and $W_{-1}(\xi)$ (lower curve) are depicted for
solution to the RBH. The degenerate event horizon at
($q_{ext},x_{ext}$) corresponds to the branch point of the Lambert
function at $\xi=-1/e$. (b) Graphs for horizons $x_+$ and $x_-$ as
the solution to $G(r)=0$. The solid line denotes the magnetically
charged RBH, while the dotted curve is for the RN black hole
($a=0$ limit). A dot ($\bullet$) represents the position of
extremal RBH which satisfies $G'(r)=0$ further.} \label{fig.1}
\end{figure}
In Fig. 1b  we have shown that the solid line denotes the
magnetically charged RBH: the upper curve describes the outer
horizon $x_+$ while the lower curve the inner horizon $x_-$,
separated by the real branches of the Lambert function. The
degenerate event horizon appears at $(q_{ext}=1.056,
x_{ext}=0.871)$. On the other hand, the dotted curve is for the
Reissner-Norstr\"om black hole where the upper curve  describes
the outer horizon, while the lower curve the inner horizon. These
are coalesced into the extremal point at
$(q_{ext},x_{ext})=(1,1)$, which is different from that of the
nonlinear Maxwell case of the magnetically charged RBH.

The causal structure of the RBH is similar to that of the RN black
hole, with the internal singularities replaced by regular
centers~\cite{HAY}. As is shown in Fig. 2, the Penrose diagram of
the  extremal RBH is identical  to that of the  extremal RN black
hole except replacing the wave line at $r=0$ by the solid
line~\cite{Car}.

\begin{figure}[t!]
   \centering
   \includegraphics{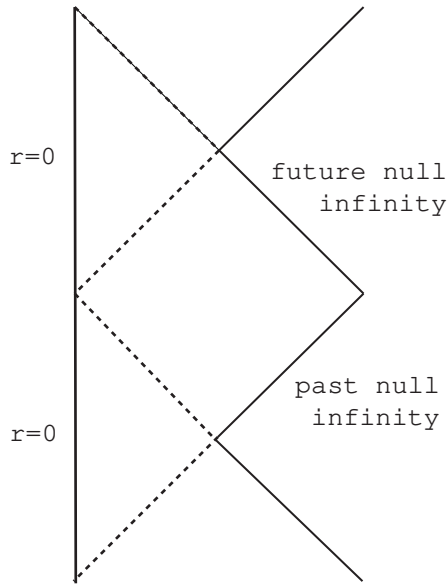}
\caption{The Penrose diagram of extremal RBH. The left (oblique)
lines denote $r=0~(r=\infty)$, while the dotted lines represent
the degenerate horizon $r=r_{ext}$.  This diagram is identical to
the extremal RN black hole except replacing the wave line at $r=0$
by the solid line. } \label{fig.2}
\end{figure}

We are in a position to investigate the near horizon geometry of
the degenerate horizon $G(r) \simeq D(r-r_{ext})^2$ defined by
$G'(r_{ext})=0$ and $G''(r_{ext})=2D$. For this purpose,  one
could introduce new coordinates $r= r_{ext} + \varepsilon/(Dy$)
and $\tilde{t}=t/\varepsilon$ with
\begin{equation}
 D=\frac{(1+\omega_o)^3}{32M^2 \omega_{o}^2}.
 \end{equation}
  Expanding the function $G(r)$ in terms of $\varepsilon$,
retaining quadratic terms and subsequently taking the limit of
$\varepsilon \to 0$, the line element~\cite{0403109} becomes
\begin{equation}
ds^2_{NH} \simeq \frac{1}{Dy^2} \left( -dt^2 + dy^2 \right) +
r^2_{ext} d\Omega_2^2.
\end{equation}
Moreover, using the Poincar\`{e} coordinate $y =1/u$, one could
rewrite the above line element as the standard form of
$AdS_2\times S^2$
\begin{equation}
ds^2_{NH} \simeq \frac{1}{D} \left( - u^2 dt^2 + \frac{1}{u^2}du^2
\right) + r^2_{ext} d\Omega_2^2.
\end{equation}
In the case of the Reissner-Nordstr\"om black hole, we have the
Bertotti-Bobinson geometry with $1/D=r_{ext}^2=Q^2$.

 For our
purpose, let us define the Bekenstein-Hawking entropy for the
magnetically charged extremal RBH
\begin{equation} \label{BHRBH}
S_{BH}= \pi r^2_{ext}=\pi M^2 x^2_{ext}=\pi Q_{ext}^2\Big[\frac{4
q_{ext}}{4+ q^2_{ext}}\Big]^2
\end{equation}
with $Q_{ext}=Mq_{ext}$. On the other hand, it is a nontrivial
task to find the higher curvature corrections to the
Bekenstein-Hawking entropy in Eq.(\ref{BHRBH}) when considering
together Einstein gravity-nonlinear electromagnetics with the
higher curvature terms~\cite{0606185}.

\section{Entropy of extremal RBH}

 Since the magnetically charged extremal RBH
is an interesting object whose near horizon geometry is given by
topology $AdS_2\times S^2$ and whose action is already known, we
attempt to obtain the black hole entropy in Eq.(\ref{BHRBH}) through
the  entropy functional approach. According to Sen's entropy
function approach, we consider an extremal black hole solution whose
near horizon geometry is given by $AdS_2\times S^2$ with the
magnetically charged configuration

\begin{eqnarray} \label{e11} && ds^2\equiv g_{\mu\nu}dx^\mu dx^\nu =
v_1\left(-r^2 dt^2+{dr^2\over r^2}\right)  +
v_2~ d \Omega^2_2,  \\
\label{e12} &&  F_{\theta\phi} = {Q} \, \sin\theta\, ,
\end{eqnarray}
where $v_i (i=1,2)$ are constants to be determined.
For this background, the nonvanishing components of the Riemann
tensor are
\begin{eqnarray} \label{e1a}
R_{\alpha\beta\gamma\delta} &= &-v_1^{-1} (g_{\alpha\gamma}
g_{\beta\delta} - g_{\alpha\delta} g_{\beta\gamma})\, , \qquad
\alpha, \beta, \gamma, \delta =r, t\, ,
\nonumber \\
R_{mnpq} &=& v_2^{-1}\,  (g_{mp} g_{nq} - g_{mq} g_{np} )\, , \qquad
m,n,p,q = \theta, \phi \, ,
\end{eqnarray}
which are  related to$AdS_{2}$ and $S^2$ sectors, respectively.
 Let us denote by
$f(v_i, Q)$ the Lagrangian density (\ref{action}) evaluated for
the near horizon geometry (\ref{e11}) and integrated over the
angular coordinates~\cite{0505122}:
\begin{equation}
\label{e2} {f}(v_i, Q) = \frac{1}{16\pi} \int d\theta\, d\phi\,
\sqrt{- g}\, \left[R\, -\,{\cal L}(B) \right]\,.
\end{equation}
Since $ R=-\frac{2}{v_1}+\frac{2}{v_2}$ and
$B=\frac{2{Q}^2}{{v_2}^2}$, we obtain
\begin{equation}
 {f}(v_i,{Q})=\frac{1}{2}v_{1}v_2\left[-\frac{1}{v_1}+\frac{1}{v_2} \, -
\frac{1}{2}{\cal L}(v_{2},{Q}) \right].
\end{equation}
Here
\begin{equation}
{\cal L}(v_{2},{Q}) = \frac{2{Q}^2}{{v_2}^2} \cosh^{-2} \left(
\frac{{Q}^2}{2\alpha \sqrt{v_2}}\right), \label{e24}
\end{equation}
which is
 a nonlinear function of $v_2$. Further, we choose the free parameter $a=Q^{3/2}/2\alpha$.
 Then, one
 could obtain  the values of $v^e_i$ at the degenerate horizon
 by extremizing $f$:
\begin{equation}
\label{e3} {\partial f \over \partial v_i} = 0\, .
\end{equation}
On the other hand, the non-trivial components of the gauge field
equation and the Bianchi identities are already given in Eqs.
(\ref{maxwell1}) and (\ref{maxwell2}), which are automatically
satisfied by the background  (\ref{e11}) and  (\ref{e12}).  It
follows that the constant $Q$ appearing in (\ref{e12}) corresponds
to a magnetic charge of the black hole. For fixed $Q$, Eq.
(\ref{e3}) provides a set of equations, which are equal in number to
the number of unknowns $v_i$.  Hereafter we choose the free
parameter $a=Q^{3/2}/2M(\alpha=M)$ to meet the condition that the
near horizon geometry of Eq.(\ref{e11}) reflects that of the
magnetically charged extremal RBH.  For the magnetically charged
extremal RBH, the entropy function is given by
\begin{equation} \label{e31}
{\cal F}(v_i, {Q}) = -2\pi {f}(v_i, {Q}) \, .
 \end{equation}
In this case, the extremal values $v^e_i$ may be  determined by
extremizing the function ${\cal F}(v_i, {Q}) $ with respect to
$v_i$:
\begin{eqnarray}
\label{e33} {\partial {\cal F} \over \partial v_1} &=&0 \to
\frac{v_2}{2} {\cal L}(v_{2}, Q)=1\,~{\rm with}~{\cal
L}(v_{2},{Q}) = \frac{2{Q}^2}{{v_2}^2} \cosh^{-2} \left(
\frac{{Q}^2}{2M \sqrt{v_2}}\right), \\
{\p {\cal F} \over \p v_2 } &=&0 \to
\frac{1}{v_1}=\frac{Q^2}{v^2_2}\cosh^{-2}\left[\frac{Q^2}{2M\sqrt{v_2}}\right]
 -\frac{Q^2}{v_2}\frac{\partial}{\partial
v_2}\left(\cosh^{-2}\left[\frac{Q^2}{2M\sqrt{v_2}}\right]\right)\,
\label{e332}
\end{eqnarray}
which are two attractor equations.
\begin{figure}[t!]
   \centering
   \includegraphics{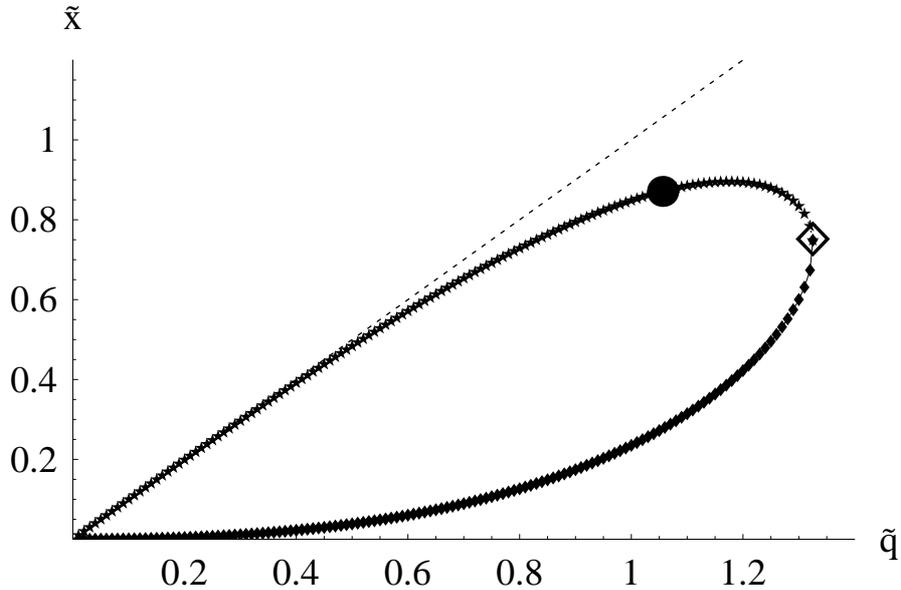}
\caption{Plot of curvature radius $\tilde{x}$ of $S^2$ versus
parameter $\tilde{q}$. The solid curve with the upper and lower
branches  denotes  the solution space to  Eq. (\ref{sol1}), while
the dotted line represents $\tilde{x}=\tilde{q}$  for the extremal
RN black hole with $v^e_2=Q^2$. A dot ($\bullet$) represents the
extremal black hole, whose conditions are given by both $G(r)=0$
and $G'(r)=0$. Diamond($\diamond$) denotes the point of
$(\tilde{q}_c,\tilde{x}_c)=(1.325,0.735)$ at which the upper and
lower branches merge.} \label{fig.3}
\end{figure}
Using the above relations,  the entropy function at the extremum
is given by
\begin{equation}
{\cal F}(v^e_{2}, Q)=\pi v^e_2.
 \end{equation}
In order to find the proper extremal value of $v^e_2$, we
introduce $Q = \tilde{q} M$, $v^e_2 = M^2 \tilde{x}^2 $ and
$v^e_1=M^2\tilde{v}_1$. Then Eqs. (\ref{e33}) and (\ref{e332})
with Eq. (\ref{e24}) can be rewritten as
\begin{eqnarray}
\label{sol1}\frac{\tilde{x}^2}{\tilde{q}^2}&=&\cosh^{-2}(\frac{\tilde{q}^2}{2\tilde{x}}), \\
\label{sol2}\frac{1}{\tilde{v}_1}&=&
\frac{\tilde{q}^2}{{\tilde{x}}^4}
\cosh^{-2}(\tilde{q}^2/2\tilde{x})
-\frac{\tilde{q}^4}{2\tilde{x}^5}
\frac{\sinh(\tilde{q}^2/2\tilde{x})}{\cosh^{3}(\tilde{q}^2/2\tilde{x})},
\end{eqnarray}
where we use $\tilde{x}$ and $\tilde{q}$ to distinguish $x$ and
$q$ for  the full equations. Note that these equations are
identical to those in Ref.{~\cite{0403109}} derived from the near
horizon geometry of an extremal RBH. This means that the entropy
function approach is equivalent to solving the Einstein equation
on the $AdS_2 \times S^2$ background, but not the full equations.

Since the above coupled equations are nonlinear equations, we
could not solve them analytically.  Instead, let us numerically
solve the nonlinear equation (\ref{sol1}) whose solutions are
depicted in Fig. 3. It seems that there are  two branches: the
upper and lower ones which merge at
$(\tilde{q}_c,\tilde{x}_c)=(1.325,0.735)$.  Note that  the
magnetically charged extremal RBH corresponds to the point
$(\tilde{q}_{ext},\tilde{x}_{ext})=(1.056, 0.871)$. However, there
is no way to fix  this point although the solution space comprises
such a point. Hence it seems that the entropy function approach
could not explicitly determine the position of $v^e_1=1/D$ and
$v^e_2=r^2_{ext}=M^2x^2_{ext}$ of the extremal RBH. We note the
case of $G(r)=0,G'(r)=0\to r=r_{ext},Q=Q_{ext}$, which implies
$\frac{v^e_2}{2}{\cal L}(v^e_{2}, Q_{ext})=1$ as dot ($\bullet$)
in Fig. 3. On the other hand, the  case of $\frac{v_2}{2} {\cal
L}(v_{2}, Q)=1$ does not lead to the extremal point.

As a result, the entropy function does not lead to the
Bekenstein-Hawking entropy (\ref{BHRBH}) for the case of the
magnetically charged extremal RBH as follows:
\begin{equation}
{\cal F}= \pi v^e_2= \pi M^2 \tilde{x}^2 \neq S_{BH}=\pi
M^2x_{ext}^2.
\end{equation}
Therefore, we do not need to consider the higher curvature
corrections because the entropy function approach does not work even
at the level of $R$-gravity.

 \begin{figure}[t!]
   \centering
   \includegraphics{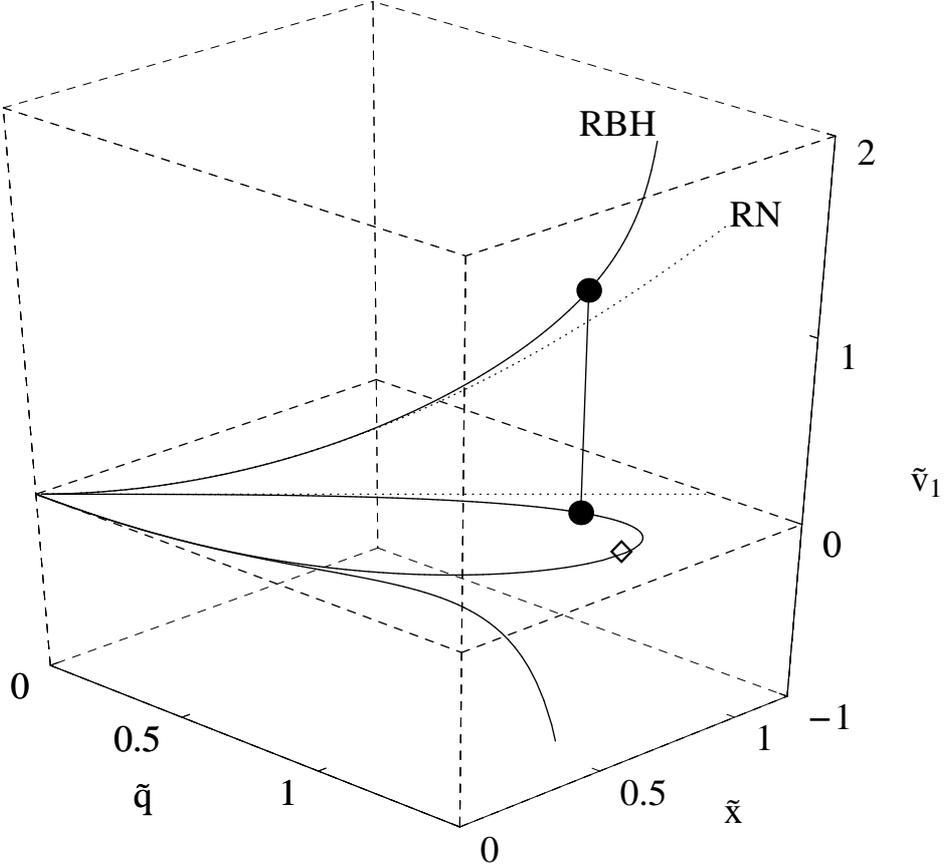}
\caption{Figure of $\tilde{v}_1$ as a function of $\tilde{q}$ and
$\tilde{x}$. Solid curve $\tilde{v}_1$, which is a monotonically
increasing function of $\tilde{q}$ and $\tilde{x}$ for the upper
branch, denotes the solution space to Eq. (\ref{sol2}) (attractor
equation). The lower branch takes negative values and thus it is
ruled out from the solution space. Dotted curve shows that of
extremal RN black hole. A dot ($\bullet$) represents the extremal
RBH.} \label{fig.4}
\end{figure}

At this stage, it seems appropriate  to comment on the case of the
singular Reissner-Nordstr\"om black hole on the $AdS_2 \times S^2$
background ($a=0$ limit). In this case, we have the entropy
function as
\begin{equation}
{\cal F}^{RN}(v_i, Q) =\pi \Big[ v_2-v_1 +  Q^2
\frac{v_1}{v_2}\Big].
 \end{equation}
Considering the extremizing process of $\p {\cal F}^{RN}/\p v_i=0$,
we find $v^{e}_2= Q^2 =v^{e}_1$, which determines the near horizon
geometry of the extremal Reissner-Nordstr\"om black hole completely.
Then, we obtain the entropy function
\begin{equation}
{\cal F}^{RN}= \pi v^{e}_2= \pi Q^2=S^{RN}_{BH},
 \end{equation}
which shows that the entropy function approach exactly reproduces
the Bekenstein-Hawking entropy, in contrast to the case of the
magnetically charged extremal RBH. Note that introducing
$v^{e}_2=M^2 \tilde{x}^2$, $Q=M \tilde{q}$ and
$v^e_1=M^2\tilde{v}_1$, one obtains the relation of
$\tilde{x}=\tilde{q}$ from $v^{e}_2=Q^2$. Furthermore, the
straight line in Fig. 3 shows the extremal Reissner-Norstr\"om
black hole solution.

In order to find more information from Eq. (\ref{sol2}), let us
solve it numerically  for given ($\tilde{q},\tilde{x}$). The
corresponding three dimensional graph is shown in Fig. 4. It shows
that for the upper branch, $\tilde{v}_1$ is a monotonically
increasing function of $\tilde{q}$ and $\tilde{x}$ and thus the
extremal point $(\tilde{q}_{ext},\tilde{x}_{ext},1/D)$ is nothing
special. Since the lower branch takes negative value of
$1/\tilde{v}_1$, it does not belong to the real solution space.
Therefore, we could not find the regular extremal point
($1.056,0.871,1.188)$ with $M=1$ even for including the curvature
radius $1/\tilde{v}_1$ of $AdS_{2}$-sector. On the other hand, for
the singular case of the Reissner-Nordstr\"om black hole, the
corresponding relation is given by
$\tilde{v}_1=(\tilde{q}^2+\tilde{x}^2)/2$.

Finally, we would like to mention how to derive the
Bekenstein-Hawking entropy of the extremal RBH from the generalized
entropy formula based on the Wald's Noether charge
formalism~\cite{CC}. According to this approach, the entropy formula
takes the form
\begin{equation} \label{gwef}
S_{BH}=\frac{4 \pi }{G''(r_{ext})}\left(q e - F(r_{ext})\right),
\end{equation}
where the generalized entropy function $F$ is given by
\begin{equation}
F(r_{ext})= \frac{1}{16\pi}\int_{r=r_{ext}} d\theta d\varphi ~r^2
\left[R-{\cal L}_M(r,Q)\right]
\end{equation}
with the curvature scalar  and the matter
\begin{eqnarray}
&& R=-\frac{r^2G''+4 r G'+2G -2}{r^2},\nonumber\\
 && {\cal L}_M(r,Q)=\frac{2Q^2}{r^4}\cosh^{-2}\left[\frac{Q^2}{2Mr}\right].
\end{eqnarray}
In this approach, one has to know the location $r=r_{ext}$ of
degenerate event horizon (solution to full Einstein equation:
$G(r)=0,G'(r)=0$). After the integration of angular coordinates,
the generalized entropy function leads to
\begin{equation}
F(r_{ext})= \frac{1}{4}\Big[-r^2G''(r)+2-r^2{\cal
L}_M(r,Q)\Big]\mid_{r=r_{ext}}=-\frac{1}{4}G''(r_{ext})r_{ext}^2
\end{equation}
because of ${\cal
L}_M(r_{ext}=Mx_{ext},Q_{ext}=Mq_{ext})=2/r^2_{ext}$. For a
magnetically charged RBH with $e=0$, we have the correct form of
entropy from Eq. (\ref{gwef})
\begin{equation} \label{WEFe1}
S_{BH}=-\frac{4\pi }{G''(r_{ext})}F(r_{ext})=\pi r_{ext}^2.
\end{equation}
Even though we find the prototype of the Bekenstein-Hawking entropy
using the entropy formula based on Wald's Noether charge formalism,
there is still no way to explicitly fix the location $r=r_{ext}$ of
degenerate horizon.

\section{Discussions}
We have considered a magnetically charged  RBH  in the coupled
system of the Einstein gravity and nonlinear electrodynamics. The
black hole solution is parameterized by the ADM mass and magnetic
charge ($M,Q$), while the free parameter $a$ is adjusted to make the
resultant line element regular at the center. Here we have put
special emphasis on its extremal configuration because it has the
similar near horizon geometry $AdS_2\times S^2$ of the extremal
Reissner-Nordstr\"om black hole ($a=0$ limit). However, the near
horizon geometry of the magnetically charged extremal RBH (extremal
Reissner-Norstr\"om black hole) have  different modulus of curvature
(the same modulus). Moreover the extremal RBH is regular inside the
event horizon, whereas the extremal Reissner-Nordstr\"om black hole
is singular.

In this work, we have carefully investigated whether the entropy
function approach does also work for deriving the entropy of a
magnetically charged extremal regular black hole in the Einstein
gravity-nonlinear electrodynamics. It turns out that the entropy
function approach does not lead to a correct entropy of the
Bekenstein-Hawking entropy even at the level of $R$-gravity. This
contrasts to the case of the extremal Reissner-Nordstr\"om black
hole in the Einstein-Maxwell theory. This is mainly because the
magnetically charged extremal RBH comes from the coupled system of
the Einstein gravity and nonlinear electrodynamics with a free
parameter $a \not=0$.

It seems  that the entropy function approach is sensitive to
whether the nature of the central region of the black hole is
regular  or singular. In order to study this issue further, one
may consider another non-linear term of  the Born-Infeld action
instead of the nonlinear electrodynamics on the Maxwell-side. It
turned out that for a singular black hole with four electric
charges, the entropy function approach does not lead to the
Bekenstein-Hawking entropy~\cite{0604028}. This means that the
Einstein gravity-Born-Infeld theory do not have a nice extremal
limit when using the entropy function approach. Hence, we suggest
that the nonlinearity on the Maxwell-side makes the entropy
function approach useless in deriving the entropy of the extremal
black hole.

Furthermore, we mention the attractor mechanism. The entropy
function approach did not work because the free parameter  is
fixed to be $a=Q^{3/2}/2M$. This could be explained by the
attractor mechanism which states that the near horizon geometry of
the extremal black holes depends only on the charges carried by
the black hole and not on the other details of the
theory~\cite{0506177}. Thus the dynamics on the horizon is
decoupled from the rest of the space. The attractor mechanism
plays an important role in the entropy function approach. However,
this mechanism  is unlikely applied to computing the entropy of a
magnetically charged extremal regular black hole because the
parameter $a$ depends on both  the charge $Q$ and the asymptotic
value $M$.

We would like to  emphasize  our three figures again  because
these provide the important message to the reader. Fig. 1a and 1b
show the outer horizon $r_+=Mx_+$ and inner one  $r_-=Mx_-$, as
the solution space to $G(r)=0$. In order to find two horizons, we
need to solve the full equation (\ref{EEQ}) with the boundary
conditions at $r=0$ (regularity) and $r=\infty$ (ADM mass). The
location of degenerate horizon $(q_{ext},x_{ext})$  is determined
by requiring the further condition of $G'(r)=0$. Fig. 3 shows the
solution space to the attractor equation (\ref{e33}). This
equation is not sufficient to determine the location of degenerate
horizon, even though the solution space comprises such a
degenerate point. Fig. 4 implies that the lower branch in Fig. 3
is meaningless. Consequently, to determine the entropy of an
extremal RBH, we need to know the mechanism which translate the
full equation to determine $G(r)=0$ into the extremal process of
attractor equation.

Finally, we note that the failure of the entropy function approach
to a magnetically charged extremal RBH is mainly  due to the
nonlinearity of the matter action (\ref{lagr}) with $a\not=0$. Of
course, this nonlinear action is needed to preserve the regularity
at the origin of coordinate $r=0$. Furthermore, the regular
condition of $a=Q^{3/2}/2M$ requires an asymptotic value of the
ADM mass $M$, in addition to charge $Q$. Considering the $a=0$
limit, we find the linear action of the Einstein-Maxwell field,
where the entropy function approach works well for obtaining the
extremal RN black hole. For the nonextremal RBH, the
Bekenstein-Hawking entropy provides $S_{BH}=\pi r_+^2=\pi M^2
x_+^2$ as its entropy because the entropy function approach was
designed only for finding the entropy of  extremal black holes.

In conclusion, we have explicitly shown that the entropy function
approach does not work for a magnetically charged extremal regular
black hole, which is obtained from  the coupled system of the
Einstein gravity and nonlinear electrodynamics.

\vspace{0.5cm}

\medskip
\section*{Acknowledgments}
This work was supported by the Science Research Center Program of
the Korea Science and Engineering Foundation through the Center for
Quantum Spacetime of Sogang University with grant number
R11-2005-021. This work of Y.-W. Kim was supported by the Korea
Research Foundation Grant funded by Korea Government (MOEHRD)
(KRF-2007-359-C00007).

\end{document}